\documentstyle[prd,aps,epsfig,floats]{revtex}
\tighten
\begin{document}
\draft
\title{The estimation of nuclear fusion rates
for ${}^{17}O$${}^1H\longrightarrow {}^{18}F^*$ transition from the
bound state and electron screening effect for the scattering state.}
\author{V.~B.~Belyaev${}^1$ \and
D.~V.~Naumov${}^{1,2}$ \and and F.~M.~Pen'kov${}^{1,3}$}
\address{${}^1$ Joint Institute for Nuclear Research, Dubna, Russia \\
         ${}^2$ Physics Department, Irkutsk State University, Irkutsk
         664003, Russia \\
         ${}^3$ Physics Department, Alma-Ata State University,
         Alma-Ata, Kazakhstan \\
         }
\date{\today}
\maketitle
\begin{abstract}
We estimate the nuclear fusion rates for the transition
${}^{17}O$${}^1H \longrightarrow {}^{18}F^*$ from the ground
and excited molecular states and also consider effect of screening
of electric charges of nuclei by bound electrons in the case of
colliding atoms ${}^{17}O$ and ${}^1H$.  We show that the fusion rate
for the molecular state is vanishingly small while the electron
screening drastically changes the rate for the colliding atoms at
kinetic energies up to hundreds of keV. The latter may be of
consequence for some astrophysical applications.
\end{abstract}
\section{Introduction}
It is a matter of common knowledge that the nuclear fusion rate
from a molecular state is strongly suppressed by Coulomb potential
barrier and such fusion events are undetectable. Our interest in the
issue is provoked by potential possibility of enhancement of
nuclear fusion rates from molecular states due to degeneration of
energies of initial and final states~\cite{Belyaev}. In this
paper, it was shown that the transition amplitude
for the $d\,{}^6Li \longrightarrow {}^8Be$ reaction increases due to
degeneration of energies, $E_{d\,{}^6Li}=22.2$ MeV and
$E_{{}^8Be}=22.279$ MeV (``nuclear resonance'').  As a simple
assumption, the following resonance wave function of $d\,{}^6Li$ was
chosen:
$$
\Psi_{{\rm res}}(r)=\frac{1}{N_{{\rm res}}}\frac{e^{i\eta \ln kr}}{r},
$$
with $\eta=Z_1Z_2\alpha\sqrt{\mu c^2/2E}$. The molecular wave function was
constructed as a product of the regular Coulomb solution $F_0(k,r)$
and an exponentially decreasing function associated with size of the
molecule,
$$ \Psi_{{\rm mol}}(r)=\frac{1}{N_{{\rm mol}}}\frac{F_0(k,r)}{r}e^{-kr}.
$$
Normalization factors $N_{{\rm res}}$ and $N_{{\rm mol}}$ were not
calculated. It was found out that amplitude transition (taken as
overlap integral of $\Psi_{{\rm res}}$ and $\Psi_{{\rm mol}}$)
increases as $\eta^{3/2}e^{0.114\pi\eta}$,
instead of the usual Coulomb barrier factor $e^{-\pi\eta}$. It was
proposed that such an enhancement may even be detectable if the
normalization factors $N_{{\rm res}}$ and $N_{{\rm mol}}$ are not too
large. Apparently, this speculation may be generalized to
other relevant molecular states.
So, we decided to elucidate what is actually going on,
using a more rigorous approach to the problem.  However we do not
consider nuclear forces explicitely and associated nuclear resonance.
We restrict our calculations by electromagnetic forces emphasizing
value of wave function discribing relative motion of two nuclei in
molecule at nuclear scale distances. Effect of nuclear resonance
leading to the strong enhancement of the fusion rate could be
included later in astrophysical $S$-factor.  To be specific, we
consider the hydroxyl molecule that is the system of two nuclei,
${}^{17}O$ and ${}^1H$, with nine electrons around.  This choice is
made taking into account the degeneration between the total energy of
two nuclei ${}^{17}O$ and ${}^1H$ ($5.607$~MeV) and the energy of the
``united'' excited nucleus ${}^{18}F^*$ ($5.605$~MeV).  Solving the
relevant Hartree-Fock-Roothaan (HFR) equations for the system within
the method of molecular orbitals and taking into account the actual
distribution of electrons, we calculate the effective potential
between the two nuclei.  With this potential, the wave functions for
bound and scattering states of the nuclei are obtained. Using these
wave functions, we estimate the probability for the nuclear fusion
${}^{17}O{}^1H\longrightarrow {}^{18}F^*$ both from the molecular and
scattering states. The latter case may be of potential interest in
studies of thermonuclear reaction in stars.
For stars like the Sun, the common accepted Debye-Huckel approximation
to the calculation of screening effect is, strictly speaking irrelevant,
because of huge difference between the scattering scale (nuclear size)
and Debye radius, $R_D$ (atomic size). This fact is usually of no
importance considering that the screening due to the plasma electrons
is by itself rather small (see Ref.~\cite{Dar} for more
details).  On the other hand, the low-lying bound electrons (if
exist) do screen electric charges of nuclei much effectively.
Evidently the concentration of electro-neutral atoms inside a star is
very small.  Nevertheless, some appreciable fraction of ions with
low-lying bound electrons (which mainly define the value of the
screening) must be present in stellar plasma and therefore can affect
the common rate of thermonuclear fusions in the stars.
Obtained results for the specific case of the
${}^{17}O{}^1H\longrightarrow{}^{18}F^*$ reaction indicate very
important role of such screening for value of fusion rate and we believe
that this effect will be common for others nuclear transitions with
heavy enough ions.
\section {Calculation of the effective potential between two nuclei}
\begin{figure}
\center{\mbox{\epsfig{file=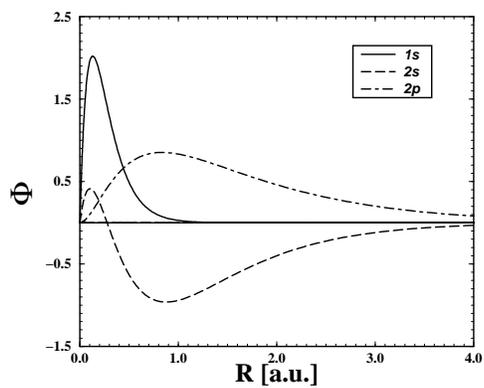,height=5.0cm}}}
        \protect\caption{Radial parts of orbital wave functions of
        oxygen}
\label{oxig}
\end{figure}

\begin{figure}
\center{\mbox{\epsfig{file=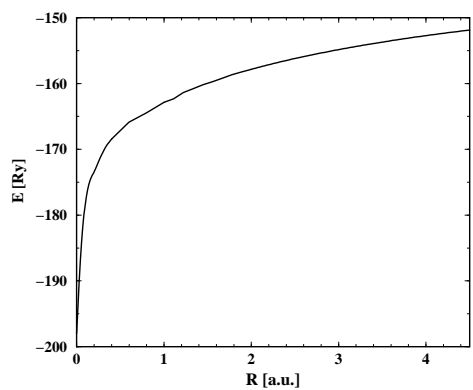,height=5.0cm}}}
        \protect\caption{Molecular energy subtracting nuclear
        repulsion energy}
\label{molenergy}
\end{figure}

Adiabatic approach allows us to split up Schr\"odinger equation for
the system into two equations both for electron and nuclear wave
functions. The approach valids up to kinetic energy of nuclei about
few MeV.
Total wave function presents as (look for instance \cite{Fudzinaga}):
$$
\Psi({\bf r}_1,\ldots,{\bf r}_9;{\bf R}_1,{\bf R}_2)=
u({\bf r}_1,\ldots,{\bf r}_9) \cdot v(R),
$$
and the equation for wave function of electrons looks:
\begin{equation}
\label{elwave}
\left[-\frac{1}{2}\sum_{i=1}^9 \triangle_i +
      \hat{V}
\right] u({\bf r}_1,\ldots,{\bf r}_9)=
E(R) \cdot u({\bf r}_1,\ldots,{\bf r}_9),
\end{equation}
where $E(R)$ depends on the $R=|{\bf R}_1-{\bf R}_2|$. Wave
function describing relative motion of nuclei ${}^{17}O$ and ${}^1H$
can be found from the equation:
\begin{equation}
\label{nuwave}
\left[-\sum_{\alpha=1}^2 \frac{1}{2M_{\alpha}} \triangle_{\alpha}
      +E(R)
\right] v(R)=\epsilon \cdot v(R),
\end{equation}
where
$$
\hat{V}=
       -\sum_{\alpha,\it i} \frac{Z_\alpha}{r_{\alpha \it i}}
       +\sum_{i > j} \frac{1}{r_{i j}}
       +\frac{Z_1 Z_2}{R}.
$$
Here $r_{\alpha \it i}$ - the distance between $\it
i$ - electron and $\alpha$ - nucleus, $r_{i j}$ - the distance
between $\it i$ and $\it j$ - electrons.
$Z_\alpha$ and $M_\alpha$ - are electric charge and mass of
$\alpha$ nucleus correspondingly measured in atomic units.
Equation (\ref{elwave}) gives us so called molecular therm $E(R)$
serving as effective potential energy for relative motion of
the nuclei. With effective potential $E(R)$ in hand we can obtain
wave function for nuclei solving (\ref{nuwave}).
At first, let's treat with $E(R)$. Wave function
$u({\bf r}_1,\ldots,{\bf r}_9)$ presents as Slater determinant whose
elements are one-electron functions called molecular spin-orbitals.
Energy of the system reaches its minimum for those of them which
satisfy to Hartree-Fock integro-differential equations which
however hard to compute for molecules.  Roothaan suggested to express
them as a linear combination of known functions $u_k(\vec r)$ with
unknown coefficients (see Ref~\cite{Fudzinaga}).  These coefficients
(formative matrix ${\bf C}$) in its turn satisfy to nonlinear
algebraic system of equations which solves iterativly:
\begin{equation}
\label{HFR}
{\bf F C = \varepsilon S C}
\end{equation}
Here ${\bf F}$ is a matrix containing matrix elements of Fock
operator ${\bf \hat F}$ dependent on internuclear distance $R$.
(see appendix A).  It depends also on unknown yet ${\bf C}$. Overlap
matrix ${\bf S}= \parallel\langle u_i|u_k\rangle\parallel$.
Diagonal matrix ${\bf \varepsilon}$ contains orbital energies as the
solution of equation (\ref{HFR}).  Configuration of molecular
spin-orbitals describing ground state of the molecule ${}^{17}O{}^1H$
is $(1\sigma){}^2(2\sigma){}^2(3\sigma){}^2(\pi^{+}){}^2\pi^{-}$
\cite{Freeman}.
So we choose molecular spin-orbitals as a linear
composition of atomic orbitals with unknown yet coefficients $C_{mi}$:
\begin{equation}
\label{configuration}
\left\{\begin{array}{ccl}
m \sigma &=& C_{m1}(1s)+C_{m2}(2S)+C_{m3}(2p_0)+C_{m4}(1h), \\
\pi{}^+ &=&(2p_+),\\
\pi{}^- &=&(2p_-),
\end{array}\right.
\end{equation}
where $m$ runs from 1 to 4.
Atomic orbitals presented here are the solution of Hartree-Fock (HF)
equations for oxygen atom \cite{footnote} and $1s$ hydrogen function:
$$
\left\{
\begin{array}{ccl}
1s       &=& (P_{1s}/r)\,Y_{00}(\theta, \varphi),   \\
2S       &=& (P_{2S}/r)\,Y_{00}(\theta, \varphi),   \\
2p_0     &=& (P_{2p}/r)\,Y_{10}(\theta, \varphi),   \\
2p_{\pm} &=& (P_{2p}/r)\,Y_{1 \pm}(\theta, \varphi),\\
1h       &=& e^{-r'}\,Y_{00}(\theta', \varphi').
\end{array}
\right.
$$
The coordinate frame centered on the oxigen atom is used.
The primed variables refer to the hydrogen nucleus.
Radial parts of these wave functions are presented in
Fig.~\ref{oxig}.
All two-centered integrals were evaluated using following presentation
for hydrogen function \cite{Freeman}:
\begin{equation}
e^{-r'}=\sum_{n=0}^{10} \frac{\alpha_n(r,R)}{r}P_n(\cos\theta).
\label{alpha}
\end{equation}
Here $r'=\sqrt{r^2+R^2-2rR\cos\theta}$,
coefficients $\alpha_n(r,R)$ -  are linear
combination of half-integer order Bessel functions (see appendix B).
$P_n(\cos\theta)$ - are Legendre polynomials. It allow us to integrate
over all angles in appearing integrals and to reduce them to one- and
two dimensioned. Calculation of the kinetic integrals requires also
derivation of $\alpha_n(r,R)$ over $r$, all these functions were
FORTRAN coded and multiply tested. Used expansion (\ref{alpha})
converges rapidly with increasing $n$.
Equation (\ref{HFR}) has a $6\times6$ block-diagonal structure for
spin-orbitals with different symmetries. Thus $\sigma$- orbitals are
the solution of $4 \times 4$ secular equation (\ref{HFR}) and $\pi$-
orbitals can be deduced from the normalization conditions.
Coinciding symmetries of the ground state of hydroxil molecule
($\pi$-symmetry) and that of the united atom Ftor (p-symmetry) for
$R\longrightarrow 0$ allow us to use the configuration of
spin-orbitals (\ref{configuration}) for all internuclear distances
from $R=0$ to $R=\infty$.  For a lot of molecules these symmetries
are different and one has to separate region of internuclear
distances for regions with different spin-orbital configurations and
then to concatenate solutions from different regions.  So in our case
we can easely check out if HFR method works well for
${}^{17}O$${}^1H$ molecule simlpy running $R$ in the equation
(\ref{HFR}).  Molecular energy subtracting nuclear repulsion energy
was obtained for $R \in (0,4)$ a.u.  with step $0.1$ a.u. and spline
interpolated.  It is presented in Fig.~\ref{molenergy}.
Calculated molecular energy smoothly approaches to that of
united atom (Ftor) ($E_F\simeq -199 Ry$) and gives good
approximation for energy of isolated atoms:  $E_{O+H} \simeq -149
Ry$. Molecular energy reaches the minimum for $R\simeq 2.2$ a.u., and
ratio of calculated energy $E_{min}=-149.86$ Ry to the experimental
value $-151.56$ Ry \cite{Freeman} at this point is $0.988$.  So in
three points where there are experimental results for the potential
curve we have a good agreement with experiment.
\section{Wave functions of the nuclei}
\subsection{Bound state}
\begin{figure}
\center{\mbox{\epsfig{file=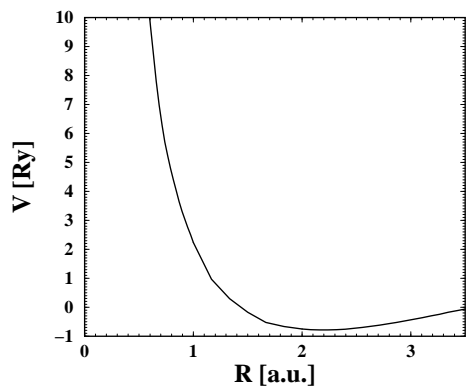,height=5.0cm}}}
        \protect\caption{Effective potential between two nuclei}
        \label{potential}
\end{figure}

\begin{figure}
\center{\mbox{\epsfig{file=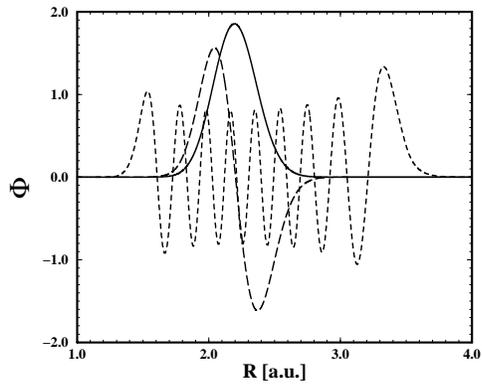,height=5.0cm}}}
        \protect\caption{Radial parts of the ground, first excited
        and 16-th excited wave functions}
\label{waves}
\end{figure}

Effective potential between the two nuclei shifted for the energy of
isolated atoms and measured in Ry is presented in
Fig.~\ref{potential}.  Numerical integration of the radial equation
of relative motion of the two nuclei with orbital moment $L=0$ was
accomplished by means of Numerov scheme.  Discrete vibrational
spectrum of nuclei was found using method of "shooting".
Table \ref{bound} presents the results. Energy counts from the bottom
of the potential curve. Vibrational energy spectrum with
$\triangle E = 0.04$ Ry $=0.54$ eV is presented here.
Value of $\Psi(0)$ is measured in
atomic units.  We also (as in \cite{Picker}) note significant
enhancement of the rates of nuclear fusion for molecular vibrational
excited states.  Physical reason for such behavior one could simply
understand looking on Fig.~\ref{waves}. There are shown
ground, first excited and 16-th excited wave functions. It is
seen that the last one is much more near to the $R=0$ point.
{\it Nonetheless vanishingly small value of $\Psi(0)$ forbids the
possibility for any cold nuclear fusion from the molecular state
${}^{17}O$${}^1H$ even for the system with degenerate energies.}
Estimation for the fusion rate gives \cite{Jackson}:
$$
\lambda=1.175 \times 10^{9} {\rm \frac{S(E)}{ MeV \cdot  barn}}\times
|\Psi(0)|^2 \cdot \mbox{sec${}^{-1}$},
$$
where $S(E)$ -is astrophysical $S$ factor. Numerical factor in this
formula is given for value of $\Psi(0)$ measured in atomic units.
\subsection{Scattering state}
\begin{figure}
\center{\mbox{\epsfig{file=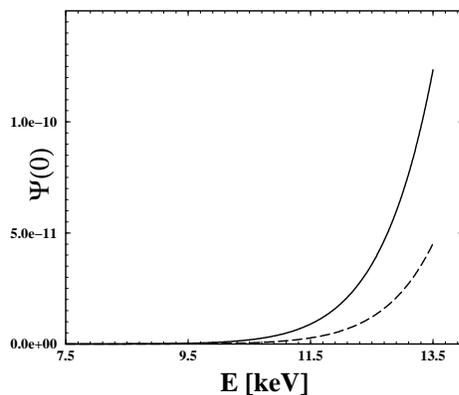,height=5.0cm}}}
        \protect \caption{Scattering wave function at $R=0$ of colliding
        nuclei with electron screening (solid line) and for
        bare nuclei (dashed line)}
\label{psi0}
\end{figure}

\begin{figure}
\center{\mbox{\epsfig{file=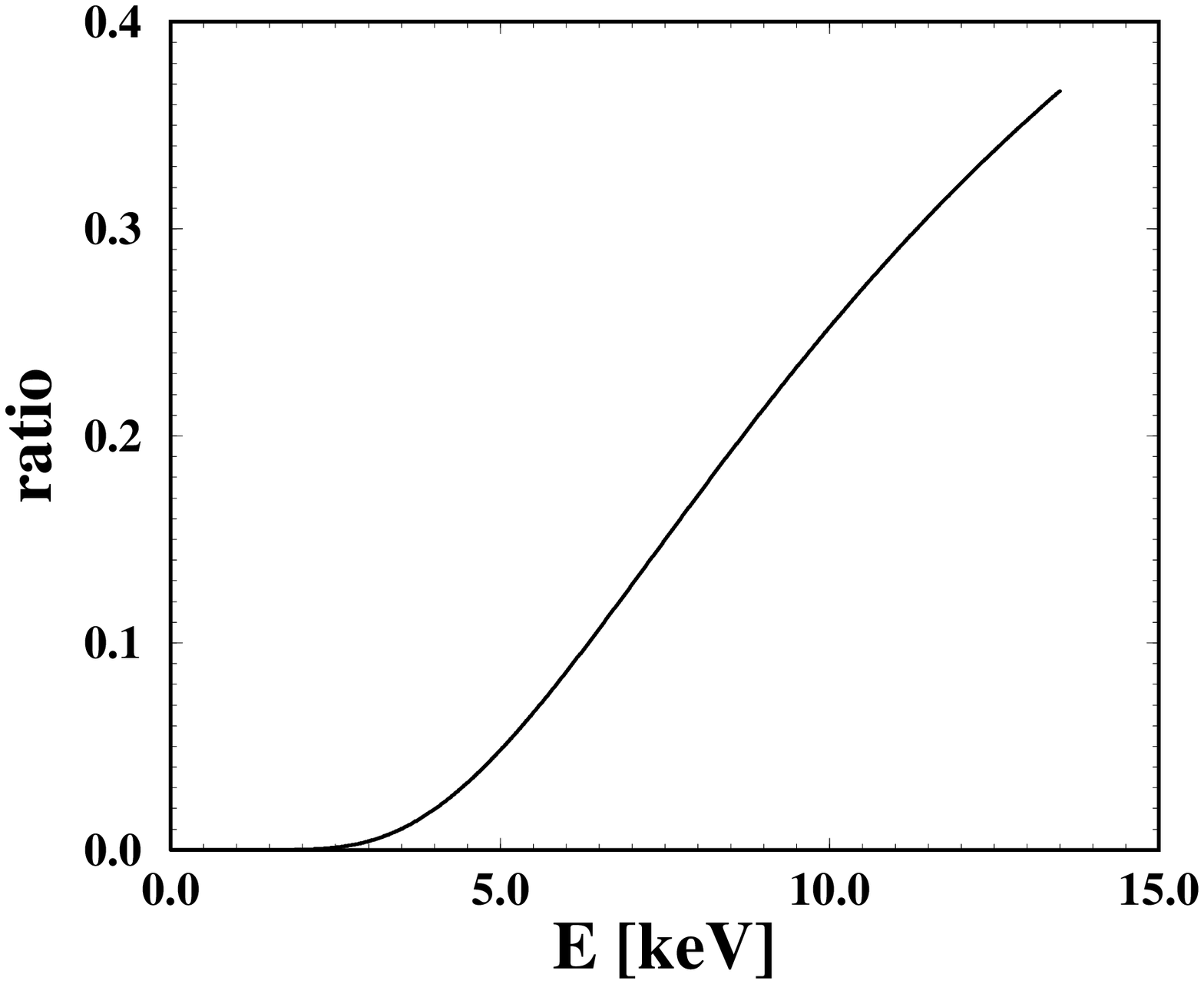,height=5.0cm}}}
        \protect\caption{ Ratio $\frac{\Psi_{\rm Coul}(0)}{\Psi(0)}$
        for different kinetic energy of the nuclei}
\label{ratio}
\end{figure}

Practically nuclear fusions from scattering state of nuclei are
much more interesting. Thermonuclear reactions in stars and nuclear
fusions in laboratories on the Earth are such examples.  Whereat
cross sections of these events depend on wherever existing electrons
which screen electric charges of nuclei. Thereby correct accounting
of electron screening of electric charges of nuclei is required.
It is a simple matter to show that common accepted Debye-Huckel
approximation to the calculation of screening effect is irrelevant
for not very dense stars like the Sun. This approach requires
electroneutral plasma, i.e. there are many charged particles within a
Debye sphere $R_D$. However most of the contribution to the nuclear
reaction rates comes from the internuclear distances shorter then
Bohr radius of colliding nuclei. In the core of the Sun scale of
$R_D$ is atomic one and scale of the the Bohr radius is
nuclear one.  Since there are no electrons inside a sphere with
nuclear size in solar plasma we come to the conclusion that
Debye-Huckel approximations becomes irrelevant for above considered
situations.  Recently this question was discussed by Dar and Shaviv
~\cite{Dar} in relation with calculation of
neutrino fluxes from the Sun.  What actually could screen effectively
electric charge of colliding nuclei are atomic electrons. To learn
effect of atomic electron screening we use effective potential
between two nuclei obtained in this work.
Assuming asymptotic wave function describing relative motion with the
momenta $p$ of the two nuclei as:
\begin{equation}
\Psi(r) \simeq \frac{sin(pr+\delta)}{r},
\end{equation}
we integrate equation (\ref{nuwave}) and compare obtained
$|\Psi(0)|^2$ with $|\Psi_{\rm Coul}(0)|^2$ that has a meaning of the
scattering wave function of the bare nuclei.
In our calculations $R_{min}=10^{-5} a.u.$ represents nuclear scale.
\begin{equation}
|\Psi_{\rm Coul}(0)|^2=e^{-2\pi\eta}2\pi \eta p^2.
\label{psiCoul}
\end{equation}
Here $\eta=m{}^*Z_1Z_2/a_0 p$, $a_0$ - is the Bohr hydrogen
radius.
Fig.~\ref{psi0} demonstrates the solution of numerical integration
of equation (\ref{nuwave}).
The ratio $\frac{\Psi_{\rm Coul}(0)}{\Psi(0)}$
is shown in Fig.~\ref{ratio}.
For kinetic energy of colliding nuclei (energies are in the
center-of-mass system) below $1$ keV, $\Psi(0)$ is many orders of
magnitude larger then $\Psi_{\rm Coul}(0)$, and  for medium kinetic
energy in the Sun ($E\simeq 1.4$ keV) the ratio
$\frac{\Psi(0)}{\Psi_{\rm Coul}(0)} \simeq 10^8$.  Since Coulomb
barrier ($\sim$ 1 MeV), most of the contribution to the fusion rates
comes from collisions with energies larger then medium energy in the
core of the Sun. For example, for $E\simeq 10$ keV, the ratio is
about $9$ and it tends to $1$ when kinetic energy of colliding atoms
reaches energy about $1$ MeV.
{\it Thereby atomic electron screening drastically
changes nuclear fusion rates and must be taken into account.}
Physical reason of such effect is "electron acceleration". It means
that the difference between energy of united atom (ftor) and energy
of separated atoms (oxygen and hydrogen) adds to the kinetic energy
of nuclei accelerating them (an extra attraction discribes by an
energy increment $U$). Thus, the nuclei may be considered as tunneling
the Coulomb barrier at an effective incident energy $E_{eff}=E+U$.
Usually, the resulting enhancement of the cross section expresses as:
\begin{equation}
\label{enhancement}
f(E)=\frac{\sigma(E+U)}{\sigma(E)}
     \simeq \exp{\{ \pi \eta(E) \frac{U}{E} \}},
\end{equation}
at those energies when $U \ll E$. For considered case of
${}^{17}O$${}^1H$ molecule, $U\simeq 648 eV$. And electron screening
due to the exponential dependence (\ref{enhancement}) at low energies
leads to the very strong enhancement of nuclear fusion rates.
Good qualitative agreement  is observed with values computed from the
formula (\ref{enhancement}) and from our calculations.
\section{conclusions}
$\bullet$
Althogh the excited state rates of molecule ${}^{17}O$${}^1H$ are many
order of magnitude larger then those for the ground states, all are
much too small to be detected even for the system with degenerate
energies.
$\bullet$
Atomic electrons do screen effectivelly electric charge of colliding
nuclei, drastically changing their fusion rates for low energies.
Although obtained results are for the specific case of the
${}^{17}O{}^1H\longrightarrow{}^{18}F^*$ fusion
we believe that this effect will be common for others
nuclear transitions with heavy enough ions.
\acknowledgments
We are grateful to R.~Cowan for help in using his FORTRAN code rcn.f
and to V.~Naumov for use of his FORTRAN code Mul.f computing
multi-dimensional integrals and for encouragement.
 \protect \appendix
\section{}

$\bullet$ Matrix element of Fock operator ${\bf F}$ used in this
work looks:

$$
F_{km} = H_{km} + \sum_{l,n=1}^6D_{ln}(2[km|ln]-[kn|lm]),
$$
where
\begin{eqnarray*}
&&D_{ln} = \sum_{i={\rm occup. orb.}}C_{i l}{}^*C_{i n}, \qquad \qquad
H_{km}=<u_k|\hat h|u_m>=
\int d{\bf r} u_k{}^*({\bf r}) \hat h({\bf r}) u_m({\bf r}), \\
&&{}[km|ln]= \int \int d{\bf r}_1 d{\bf r}_2
\frac{u_k{}^*({\bf r}_1)u_m({\bf r}_1)u_l{}^*({\bf r}_2)u_n({\bf
r}_2)}
{|{\bf r}_1-{\bf r}_2|}, \qquad
\hat h({\bf r}_i)=-\frac{1}{2}\triangle_i
                     -\sum_{\alpha}\frac{Z_{\alpha}}{r_{i\alpha}}
\end{eqnarray*}

$\bullet$ Molecular energy finds as:

$$
E=\sum_{i={\rm occup. orb.}} \varepsilon^i +
  \sum_{k,m}D_{km}H_{mk}.
$$

$\varepsilon^i$ are the diagonal elements of matrix $\varepsilon$.

\section{}

$\bullet$ All two-centered integrals were evaluated using
expression (\ref{alpha}).

Denote:

$$
f_n(z)=\sqrt{\frac{\pi}{2z}}I_{n+\frac{1}{2}}(z),\quad
g_n(z)=\frac{2}{\pi} \sqrt{\frac{\pi}{2z}}K_{n+\frac{1}{2}}(z),
$$

where $I_{n+\frac{1}{2}}(z)$ and $K_{n+\frac{1}{2}}(z)$
- are modified Bessel functions of half-integer order.

Let $r'=\sqrt{r^2+R^2-2rR\cos\theta}$, then (see for
instance \cite{Abramovic}):

$$
\frac{e^{-\lambda r'}}{\lambda r'}=\sum_{n=0}^\infty
\left[(2n+1)f_n(z_1)g_n(z_2)\right]
P_n(\cos\theta).
$$

where $z_1=\lambda r_< ,z_2=\lambda r_>$.
And $r_<,\; r_>$- are minimal and maximal distance
from $r$ and $R$ correspondingly.

One can obtain:

$$
-\frac{\alpha_n(z_1,z_2)}{r}=
(2n+1)f_n(z_1)g_n(z_2)+z_1f_{n_1}(z_1)g_n(z_2)-z_2g_{n+1}(z_2)f_n(z_1).
$$
We need also derivation of these functions over $r$ for evaluating
kinetic integrals:

for $r<R$:

\begin{eqnarray*}
-\frac{\partial \alpha_n(z_1,z_2)}{\partial r}&=&
(2n+1)(n+1)f_n(z_1)g_n(z_2)+
(3n+4)z_1f_{n+1}(z_1)g_n(z_2)- \\
&&(n+1)z_2g_{n+1}(z_2)f_n(z_1)-
z_1z_2g_{n+1}(z_2)f_{n+1}(z_1)+
z_1^2g_n(z_2)f_{n+2}(z_1).
\end{eqnarray*}

for $r>R$:
\begin{eqnarray*}
-\frac{\partial \alpha_n(z_1,z_2)}{\partial r}&=&
(2n+1)(n+1)f_n(z_2)g_n(z_1)-
(3n+4)z_1f_n(z_2)g_{n+1}(z_1)+ \\
&&(n+1)z_2g_n(z_1)f_{n+1}(z_2)-
z_1z_2g_{n+1}(z_1)f_{n+1}(z_2)+
z_1^2g_n(z_1)f_{n+2}(z_2).
\end{eqnarray*}

\begin{table*}
\protect \caption{Discrete vibrational energy spectrum and $\Psi(0)$}
\label{bound}
\center
\begin{tabular}{cccccc}
energy level,N & energy, Ry & $\Psi(0),a.u.$ &
energy level,N & energy, Ry & $\Psi(0),a.u.$ \\
\hline
$0 $ & $-0.764$ & $ 1.0 \times 10^{-113}$  &
$9 $ & $-0.401$ & $ 8.9 \times 10^{-108}$  \\
$1 $ & $-0.721$ & $ 9.6 \times 10^{-113}$  &
$10$ & $-0.360$ & $ 2.0 \times 10^{-107}$  \\
$2 $ & $-0.678$ & $ 6.5 \times 10^{-112}$  &
$11$ & $-0.319$ & $ 4.8 \times 10^{-107}$  \\
$3 $ & $-0.636$ & $ 3.7 \times 10^{-111}$  &
$12$ & $-0.277$ & $ 1.2 \times 10^{-106}$  \\
$4 $ & $-0.594$ & $ 1.8 \times 10^{-110}$  &
$13$ & $-0.236$ & $ 3.0 \times 10^{-106}$  \\
$5 $ & $-0.552$ & $ 8.8 \times 10^{-110}$  &
$14$ & $-0.195$ & $ 7.7 \times 10^{-106}$  \\
$6 $ & $-0.511$ & $ 4.0 \times 10^{-109}$  &
$15$ & $-0.154$ & $ 1.9 \times 10^{-105}$  \\
$7 $ & $-0.474$ & $ 1.5 \times 10^{-108}$  &
$16$ & $-0.114$ & $ 4.7 \times 10^{-105}$  \\
$8 $ & $-0.438$ & $ 4.0 \times 10^{-108}$  &
&&\\
\end{tabular}
\end{table*}

\clearpage

\begin{references}
\bibitem{footnote}
         these functions were obtained using FORTRAN code rcn.f
         developed by R.~Cowan. This program is available by
         ftp://t4.lanl.gov/pub/cowan/
\end{references}
\end{document}